\documentclass{article}
\usepackage{spconf,amsmath,graphicx}
\usepackage{multirow,booktabs,makecell,spconf,amsmath,graphicx,amsfonts,color,subfigure,arydshln}
\usepackage{makecell, caption}
\usepackage[fontsize=9.5pt]{fontsize}

\title{Sample-adaptive Data Augmentation with Progressive Scheduling}
\name{Hongxuan Lu, Biao Li}

\address{Beijing OPPO Telecommunications Corp., Ltd., Beijing, China}
\begin{document}
\captionsetup[table]{skip=5pt}
\maketitle

\begin{abstract}
Data augmentation is a widely adopted technique utilized to improve the robustness of automatic speech recognition (ASR). Employing a fixed data augmentation strategy for all training data is a common practice. However, it is important to note that there can be variations in factors such as background noise, speech rate, etc. among different samples within a single training batch. By using a fixed augmentation strategy, there is a risk that the model may reach a suboptimal state. In addition to the risks of employing a fixed augmentation strategy, the model's capabilities may differ across various training stages. To address these issues, this paper proposes the method of sample-adaptive data augmentation with progressive scheduling(PS-SapAug). The proposed method applies dynamic data augmentation in a two-stage training approach. It employs hybrid normalization to compute sample-specific augmentation parameters based on each sample's loss. Additionally, the probability of augmentation gradually increases throughout the training progression. Our method is evaluated on popular ASR benchmark datasets, including Aishell-1 and Librispeech-100h, achieving up to 8.13\% WER reduction on LibriSpeech-100h test-clean, 6.23\% on test-other, and 5.26\% on AISHELL-1 test set, which demonstrate the efficacy of our approach enhancing performance and minimizing errors.
\end{abstract}

\section{Introduction}
In recent years, end-to-end models have emerged as the primary paradigm for automatic speech recognition (ASR) applications. These models are commonly implemented using frameworks like Connectionist Temporal Classification (CTC) loss function \cite{2014rnnt, 2014deepspeech}, attention mechanism \cite{bahdanau2016, chorowski2015}, or joint CTC-attention \cite{kim2017}, in conjunction with various neural network structures such as LSTM \cite{Graves2013}, Transformer \cite{Dong2018}, and Conformer \cite{gulati2020conformer}. 

Training end-to-end ASR systems typically require a large amount of data to prevent overfitting. Data augmentation offers a promising approach to address this challenge by increasing the quantity of speech data using existing resources \cite{2023makemore, lowresource}. However, much of the existing work on data augmentation primarily focuses on exploring fixed methods to introduce more diverse data to the model. For instance, the widely effective SpecAugment \cite{Specaug2019} method applies a fixed number and size of masks to the temporal and frequency domains(log mel spectrogram) of the data features. 

Using a fixed data augmentation strategy can lead to excessive or insufficient augmentation. If the applied augmentation operations are too many or too few, it can have negative effects on the model training \cite{cubuk2019autoaugment}. For instance, within the same batch after applying SpecAugment, if a sample has a higher loss, it indicates that it is difficult for the model to learn from that sample, suggesting that the data augmentation applied may be too intense. Conversely, if a sample has a lower loss, it means that it is relatively easy for the model, indicating that the data augmentation applied may be too mild \cite{2021sapaugment}. In such cases, using manual fixed strategies makes it difficult to precisely control the intensity of data augmentation. On the other hand, allowing the model to automatically learn data augmentation strategies can better adapt to the varying learning difficulties of different samples \cite{smartaug, autoaugment, advertiseaug}, leading to optimal augmentation effects.


SpecAugment and SpecSub \cite{2021wenet} have achieved promising improvements in ASR by using fixed augmentation strategies. Sapaugment builds upon SpecAugment by introducing sample adaptation and augmentation policy search methods, leading to further enhancements. However, Sapaugment's calculation of augmentation policy solely relies on the ranking of sample losses \cite{2021sapaugment}, disregarding the individual differences among samples. Additionally, its strategy search method is time-consuming. Similarly, G-Augment adopts spatial search techniques to determine the sequence and parameter values for applying various data augmentation methods \cite{2023Gaugment}. The contributions of this paper can be summarized as follows:

In this paper, we introduce a novel training method called Sample Adaptive Data Augmentation with Progressive Scheduling. This method aims to improve the Character Error Rate (CER) performance in automatic speech recognition. The entire experiment is conducted using the Wenet framework with the Conformer model. Firstly, we perform pre-training using the basic SpecAug method. Secondly, we apply the proposed innovative method on top of the pre-trained model, making adjustments in two dimensions. At the micro-level within a single batch, we calculate the appropriate data augmentation intensity using hybrid normalization methods. At the macro level during training, we gradually adjust the probability of applying data augmentation based on the epoch. This progressive scheduling allows for adaptive data augmentation without increasing model parameters or introducing time-consuming processes. The proposed method demonstrates superior performance compared to WeNet Conformer with SpecAugment, with a relative improvement of 8.13\% on the test-clean set and 6.2\% on the test-other set in the LibriSpeech dataset. Additionally, the proposed method achieves a relative reduction of 5.2\% in Character Error Rate (CER) on the AISHELL-1 dataset.


\begin{itemize}
    \item Introduces Hybrid Normalization, a novel algorithm for calculating adaptive strategy values in this study.
    \item Progressive Learning adapts data augmentation probability based on training epochs, accounting for the model's evolving capabilities.
    \item Pre-training and adaptation involve further training a pre-trained model using the aforementioned strategy. This approach effectively utilizes available data to achieve additional performance improvements.
\end{itemize}


\section{Related Work}
In the field of speech recognition, several spectral augmentation methods have been proposed with the aim of improving the performance of speech recognition systems. These methods introduce variations in the audio spectrogram, allowing the models to learn from a more diverse set of training data.

\noindent\textbf{SpecAug} (Spectral Augmentation) \cite{Specaug2019}, 
the method applies time masking to mask blocks of time steps, and frequency masking to mask blocks of frequency steps. 
In the training process of ASR, $X\left(t, f \right)$  represents the input feature in the form of a spectrogram before data augmentation, where t represents the time dimension and f represents the frequency dimension.
\begin{itemize}
    \item Time masking technique randomly erases portions of the audio signal in the time domain. It performs $N_{t}$ rounds masking operations of the 
    $X( t_{1}:t_{2},\ f) = 0,where\ t_{1}<t_{2} < T_{max} $, resulting in a new sample $X_{t\_mask}$.
    
    \item Frequency masking technique randomly removes value in the frequency domain, it performs $N_{f}$ times operations of the 
    $X( t,\ f_{1}:f_{2}) = 0, where\ f_{1}<f_{2} < F_{max} $, resulting in a new sample $X_{f\_mask}$.
\end{itemize}

\noindent\textbf{SpecSub} (Spectral Substitution) \cite{2021wenet} is a spectral augmentation method that randomly substitutes a framed chunk with the previous chunk. 
The primary technology of SpecSub is time substitution.

\begin{itemize}
    \item Time Sub technique randomly substitutes a framed chunk with the previous chunk in the temporal dimension. It performs $N_{s}$ substitution operations of the 
    $X( t: t+\Delta_{t},\ f) = X( \mathopen{t}^{\prime}: \mathopen{t}^{\prime}+\Delta_{t},\ f) ,where\ t, \mathopen{t}^{\prime} < T_{max} -\Delta_{t} $ , resulting in a new sample $X_{t\_sub}$.
\end{itemize}

\noindent\textbf{Sapaugment} \cite{2021sapaugment} proposed a data-augmentation framework that automatically adapts the strength of the perturbation based on the loss values of the samples.
The sample sample-adaptive policy, $f$, maps the training loss value, $l$, of a sample to a scalar, $\lambda \in [0,1]$. 
The $\lambda = f(l)$ determines the amount of augmentation applied. The policy, $f$, is parameterized by two hyperparameters,$s$, $a$, and is defined using incomplete beta function (IBF) \cite{incompletBeta} as follows:
\begin{equation}    
    f_{s,a}(l)= 1 - IBF(s(1-a), s \cdot a; \l_{rank}/ B)
\end{equation}
where \(l_{rank}\) = 1,2,...,\(B\) is the ranking of the loss value \(l\) in a mini-batch, \(B\) is the mini-batch size. The range of normalized loss ranking \(l_{rank} /B\) to standardize the input to
the policy.

\begin{figure}
\centering
\includegraphics[width=8cm]{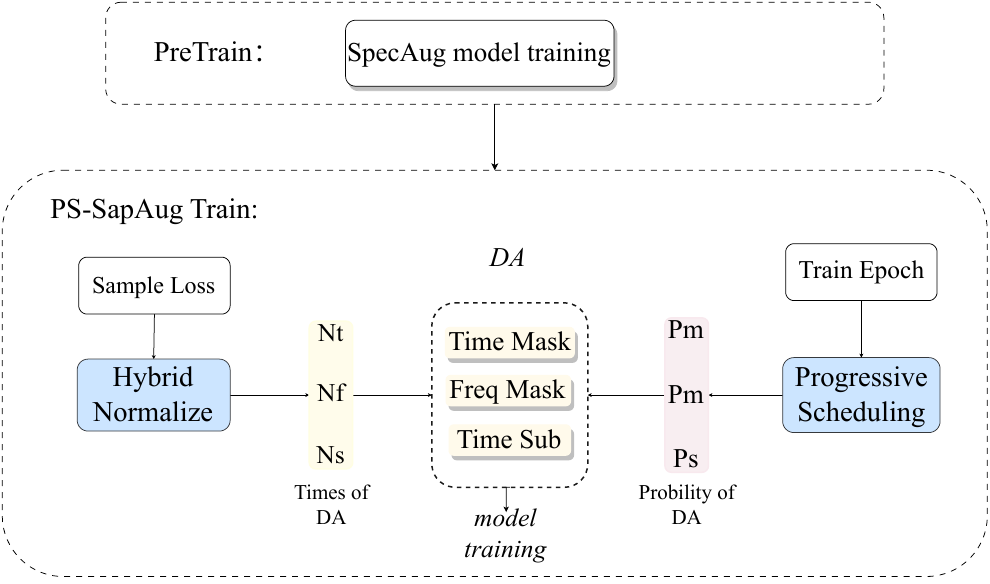}
\caption{The overall process}
\label{fig:label}
\end{figure}

\section{PROPOSED METHOD}

SpecAugment utilizes a fixed augmentation strategy without adaptive adjustments. It applies a constant number of masks, typically set to 2 in the case of WeNet Conformer \cite{2021wenet}. SapAugment \cite{2021sapaugment}introduces adaptive policy adjustments to data augmentation. However, the policy based on loss rank fails to consider the variability of individual samples within a batch. 
To address these issues, as shown in Figure 1, we propose a novel normalization calculation algorithm that is applied within a single batch to obtain sample-specific adaptive data augmentation intensity values. Additionally, in the overall training process, we progressively increase the probability of applying stronger data augmentation as the number of epochs increases.

\subsection{Hybrid Normalization policy}
In training, we begin by calculating the sample loss within a batch that hasn't experienced data augmentation. $L_{i}$ represents the loss of the i-th sample, while $L_{var}$  and $L_{mean}$ represent the variance and mean of the sample losses, respectively. The process and functions of the Hybrid Normal policy can be summarized as follows:

\begin{equation}
\mathopen{L}^{\prime}_{i} = 
\begin{cases}
  L_{mean} - 2L_{var} & \text{if } L_{i} < L_{mean} - 2L_{var} \\
  L_{mean} + 2L_{var} & \text{if } L_{i} > L_{mean} + 2L_{var} \\
  L_{i} & \text{otherwise}
\end{cases}
\end{equation}
\begin{equation}
\mathopen{L}^{\prime\prime}_{i} = \frac{\mathopen{L}^{\prime}_{i}}{\mathopen{L}^{\prime}_{i} + mean(\mathopen{L}^{\prime})}
\end{equation}
\begin{equation}
\mathopen{L}^{\prime\prime\prime}_{i} = \frac{\mathopen{L}^{\prime\prime}_{i} - \min(\mathopen{L}^{\prime\prime})}{\max(\mathopen{L}^{\prime\prime}) - \min(\mathopen{L}^{\prime\prime})}
\end{equation}
\begin{equation}
    \lambda_{i} = 1 - IBF(\mathopen{L}^{\prime\prime\prime}_{i})
\end{equation}

As shown in Equations 2, 3, and 4, the first step is to limit the noisy points (either too large or small) in the original loss values. In the second step, the processed loss values are normalized using mean normalization, which involves centering each value around the mean. Subsequently, maximum-minimum normalization is applied to scale the normalized losses within a specified range. Finally, the IBF function is employed to determine the policy $\lambda_{i}$ for each sample. 
Table 1 shows how DA is applied during training, The sample's policy $\lambda$ is mapped to the times of use of data augmentation operations using a straightforward 1-dimensional linear mapping.

\begin{table}[h]\footnotesize
\vspace{-0.2cm}
\centering
\caption{The relationship between the times of use of the three spectrum data augmentation (DA) methods and sample's Hybrid Normalization policy value $\lambda$.}
\renewcommand{\arraystretch}{1.5}
\setlength{\tabcolsep}{1mm}{
    \begin{tabular}{c | c | l | c}
    \hline
    Times of DA & Prob of DA & Mapping from $\lambda$ & Range \\ 
    \hline
    \multirow{2}{*}{$N_{t}(time\ mask)$} & $p_{mask}$ & $N_{t} = \lceil {4*\lambda} \rceil$ & $[0,1,2,3,4,5,6]$ \\
    & $1-p_{mask}$ & $N_{t} = 2$  & $[2]$ \\
    \hline
    \multirow{2}{*}{$N_{f}(freq\ mask)$} & $p_{mask}$ & $N_{f} = \lceil {4*\lambda} \rceil$ & $[0,1,2,3,4,5,6]$ \\
    & $1-p_{mask}$ & $N_{f} = 2 $ & $[2]$\\
    \hline
    \multirow{2}{*}{$N_{s}(time\ sub)$} & $p_{sub}$ & $N_{s} = \lceil {2*\lambda} \rceil$ & $[0,1,2,3,4]$ \\
    & $1-p_{sub}$ & $N_{s} = 1$ & $[1]$\\
    \hline
    \end{tabular}
}
\vspace{-0.2cm}
\end{table}

\subsection{Progressive Scheduling}
Throughout the entire training process, we have observed the following trends: In the early stages of training, the model exhibits limited capability, resulting in relatively high loss values for the training samples. However, as the model's performance improves through training, the overall loss of the entire train dataset tends to gradually decrease per epoch. This indicates that the model has successfully adapted to the existing level of data augmentation strength and adopted stronger data augmentation to train the model, enhancing its capabilities. In our work, each data augmentation method has an associated probability P to control its application. With probability P, adaptive data augmentation is performed during training. With probability 1-P, non-adaptive augmentation is performed. We use progressive scheduling to increase adaptive data augmentation strength by raising the probability P over time. The detailed process is outlined as follows:

\begin{equation}
    epoch_{policy} = IBF(epoch/total_{epoch})
\end{equation}
\begin{equation}
    p_{mask}, p_{sub} = F_{linear}(epoch_{policy})
\end{equation}


As defined by Equations 6 and 7, we first calculate the scheduling policy value $epoch_{policy}$ based on the current training epoch. Secondly, based on the $epoch_{policy}$, compute the $p_{mask}$ and $p_{sub}$ values by a linear transform function $F_{linear}$, in which the output DA probability increases as the input $epoch_{policy}$ increases. $p_{mask}$  is applied to $time\ mask$ and $freq\ mask$, and $p_{sub}$ is applied to $time\ sub$. As depicted by columns 2, 3, and 4 in Table 1, $p_{mask}$ and $p_{sub}$ determine which type of DA training method to apply. 
In Table 1, the upper halves of rows 2 through 4 present results for the adaptive method, while the lower halves show the fixed method. The adaptive technique employed more intense DA compared to the fixed approach
Consequently, $p_{mask}$ and $p_{sub}$ gradually increase along with more training epochs, indicating that the model will be trained with stronger DA as epochs progress.

\subsection{Two-stage Training}
We employ a two-stage training approach. In the first stage, we train a base model using a fixed data augmentation strategy. This pre-trained model has enough capability to recognize general samples. In the subsequent adaptive training stage, We focus on strengthening the model's performance in a more specific way.
For relatively difficult training samples, we apply fewer data augmentation techniques to allow the model to first learn their fundamental features. Once the model's capabilities adequately cover these fundamental features, we gradually increase the intensity of data augmentation. Conversely, for simpler samples, we utilize more extensive data augmentation to help the model further learn the fine-grained features of those samples.

By dividing the training process into separate stages, we can more effectively address different sample complexities. This staged approach can achieve better results.

\section{Experiments}

\subsection{Dataset}
For the speech recognition task, we utilize two datasets to evaluate this method: AISHELL-1 \cite{bu2017aishell} and LibriSpeech \cite{panayotov2015librispeech}. AISHELL-1 contains 120,098 utterances for training, 14,326 utterances for development, and 7,176 utterances for testing. LibriSpeech consists of around 1,000 hours of English audiobook speech recordings. For the purpose of efficient model training, we use the 100-hour LibriSpeech-100h subset containing US English accents. Performance is evaluated on the test-clean and test-other sets from LibriSpeech.

\subsection{Experimental settings}

\noindent\textbf{Aishell-1.}  We conducted an evaluation of the Aishell-1 dataset using the encoder-decoder architecture based on Wenet \cite{2021wenet} Conformer \cite{gulati2020conformer}. The model architecture consists of a 12-layer encoder with a feed-forward dimension of 2048 and depth-wise convolution with a kernel size of 3. The decoder comprises 6 layers with 2048 units and 4 attention heads with a dimension of 512. The acoustic feature used is an 80-dimensional Fbank computed with a 25-ms window and a 10-ms shift. During the training phase, we used the following configurations: a learning rate of 0.002, a batch size of 18, 4 GPUs were utilized, the dither was set to 0.1, and the CTC weight was set to 0.3. The model was trained for a total of 240 epochs. In the decoding phase, the CTC weight was set to 0.5.

\noindent\textbf{LibriSpeech 100h.}  For the evaluation of the LibriSpeech dataset, we employed the Wenet  Conformer architecture, using the same model configuration as Aishell-1. The training was performed with a batch size of 24 and trained for 120 epochs, utilizing a learning rate of 0.004.
80-dimensional mel-filter bank features were used for acoustic feature extraction.

\subsection{Evaluation}

In this section, we present experimental results on the Aishell-1 and LibriSpeech 100h benchmark datasets. Our proposed method, PS-SapAug, is compared against other speech data augmentation techniques, namely NoAug, SpecAug and SapAug. Additionally, pre-trained model results derived from intermediate SpecAug training are reported. The PS-SapAug results were obtained by further training the pre-trained SpecAug model.

\begin{table}[h]
\centering
\caption{Comparison of Proposed Augmentation against baselines on the Aishell-1 and Librispeech 100h dataset. All numbers are percentage character word error (CER) or word error rate (WER) (lower is better).}
\renewcommand{\arraystretch}{1.5}
    \setlength{\tabcolsep}{2mm}{
        \begin{tabular}{lc|cc}
        \toprule
        \multirow{1}{*}{\centering\textbf{Method}} & \multicolumn{1}{c}{\textbf{Aishell-1}} & \multicolumn{2}{c}{\textbf{LibriSpeech 100h}}  \\
        \cmidrule(l){2-4}
        \multirow{1}{*}{\centering{CER/WER}} & Test & Test-clean & Test-other \\
        \hline
		NoAug &5.04 &9.20  &26.82 \\
		SpecAug \cite{Specaug2019} &4.62 &8.36  &23.34 \\
		SapAugment \cite{2021sapaugment} &4.60 &8.03  &22.74 \\
		\hline
            Pre-trained &6.01 &10.66 &28.43 \\
		\hdashline
		\textbf{PS-SapAug} &\textbf{4.36} &\textbf{7.59} &\textbf{21.46} \\
		\bottomrule
        \end{tabular}
        }
\vspace{-0.2cm}
\label{tbl:overall}
\end{table}

According to the results in Table 2, our proposed PS-SapAug method achieves the best performance among the tested data augmentation methods. The pre-trained model, which came from an intermediate stage of SpecAug training, performs more poorly. These results demonstrate that PS-SapAug, compared to SapAugment utilizes a loss rank strategy and is shown to be superior.

\noindent\textbf{Aishell-1.} We compare the proposed method to the model without any augmentation, and the model trained with SpecAugment  \cite{Specaug2019} in Wenet Conformer architecture, additionally, we present the results of SapAugment for comparison. As shown in Table 2, The proposed method performs better than SapAugment and outperforms SpecAugment with up to a 5.62\% relative reduction in CER when evaluated on the Aishell-1 Mandarin speech dataset.

\noindent\textbf{Librispeech 100h.} We further evaluate our method on the Librispeech English dataset. The results demonstrate comparable performance improvements over SpecAugment and SapAugment, consistent with the result on the Aishell-1 Mandarin dataset. The proposed method demonstrates relative WER improvements of 8.13\% and 6.23\% on the Librispeech test-clean and test-other datasets respectively, compared to the baseline SpecAugment method.

\subsection{Ablation Study}

In this section, we perform ablation studies on the LibriSpeech 100h and AISHELL-1 datasets to analyze the individual contributions of each component of the proposed method. 

\begin{table}[h]\footnotesize
\fontsize{9}{10}\selectfont
\vspace{-0.2cm}
\centering
\caption{The ablation study analyzes the effect of hybrid normalization and progressive scheduling strategies on model performance. Character error rate(CER) on the Aishell-1 and Word Error Rate (WER) on the LibriSpeech 100h}
\renewcommand{\arraystretch}{1.5}
\setlength{\tabcolsep}{1.5mm}{
    \begin{tabular}{lccc}
    \toprule
        \multirow{1}{*}{\centering\textbf{Method}} &{\centering\textbf{Aishell-1}}  & \multicolumn{2}{c}{\textbf{LibriSpeech 100h}} \\
        \cmidrule(l){2-4}
        \multirow{1}{*}{\centering{WER}} &Test &Test-clean  &Test-other \\ \hline
        SpecAug  &4.62  &8.36  &23.34  \\
        \hspace{0.2cm} + Hybrid Normalization &4.50 &7.87  &22.55 \\
        \hspace{0.2cm} + Progressive Scheduling  &4.45 &7.75  &22.14 \\
        \hspace{0.7cm} \textbf{PS-SapAug} &\textbf{4.36} &\textbf{7.59} &\textbf{21.46}\\
    \bottomrule
    \end{tabular}
}
\vspace{-0.2cm}
\end{table}

As can be seen from Table 3, we conducted a study on the effects of hybrid normalization and progressive scheduling method on the experimental results. Row 1 shows the results using basic SpecAugment data augmentation as the baseline model. Row 2 shows the results by only adding our proposed hybrid normalization approach on top of the baseline model. Row 3 shows the results by only adding our proposed progressive learning rate scheduling strategy on top of the baseline model. Row 4 shows the results by adding both hybrid normalization and progressive learning rate scheduling strategies on top of the baseline model.
Based on the obtained results, we can draw the following conclusions: First, augmentation with the proposed Hybrid Normalization policy and Progressive scheduling augmentation probability can help to improve the performance compared to the SpecAugment, respectively. Second, adopting Sample Adaptive Data Augmentation with Progressive Scheduling to train is better than either one of them.


\begin{table}[h]\footnotesize
\vspace{-0.2cm}
\centering
\caption{The ablation study examines the influence of a two-stage training strategy on model performance. Character error rate(CER) on the Aishell-1 and Word Error Rate (WER) on the LibriSpeech 100h}
\renewcommand{\arraystretch}{1.5}
\fontsize{9}{10}\selectfont
\setlength{\tabcolsep}{1mm}{
    \begin{tabular}{lcc|c}
        \hline
        \multirow{1}{*}{\textbf{Method/Dataset}} & \multicolumn{1}{c}{\textbf{Aishell-1}} & \multicolumn{2}{c}{\textbf{LibriSpeech 100h}}   \\
        \cmidrule(l){2-4}
        CER/WER &Test &Test-clean  & Test-other  \\ \hline
        Standard PS-SapAug  &4.64 &8.13  &23.12  \\
        \textbf{Two-Stage} PS-SapAug & \textbf{4.36} & \textbf{7.59} & \textbf{21.46}  \\
        \bottomrule
    \end{tabular}
}
\vspace{-0.2cm}
\end{table}
As shown in Table 4, "Standard" refers to training PS-SapAug from scratch, whereas "Two-Stage" first involves SpecAug pre-training before applying PS-SapAug. When considering the Aishell-1 and LibriSpeech 100h datasets, the Two-Stage process led to markedly stronger results than the Standard approach. The results suggest that a two-stage training workflow is a more effective strategy.

\section{CONCLUSION}
In this paper, we propose a novel two-stage training approach called PS-SapAug for applying data augmentation to ASR. At the micro-level, hybrid normalization is introduced to compute sample-specific augmentation based on loss, allowing adaptive adjustment of augmentation strength. At the macro level, progressive scheduling automatically increases the augmentation probability over training to address varying model capabilities. Evaluation on Aishell-1 and Librispeech demonstrates PS-SapAug achieves significantly improved performance compared to standard fixed augmentation, highlighting the benefits of the proposed dynamic sample-adaptive and stage-aware augmentation techniques.


\end{document}